# Cluster Development and Knowledge Exchange in Supply Chain

P. Sureephong, N. Chakpitak, L. Buzon, and A. Bouras

*Abstract*— Industry cluster and supply chain are in focus of every countries which rely on knowledge-based economy. Both focus on improving the competitiveness of firm in the industry in the different aspect. This paper tries to illustrate how the industry cluster can increase the supply chain performance. Then, the proposed methodology concentrates on the collaboration and knowledge exchange in supply chain. For improving the capability of the proposed methodology, information technology is applied to facilitate the communication and the exchange of knowledge between the actors of the supply chain within the cluster. The supply chain of French stool producer was used as a case study to validate the methodology and to demonstrate the result of the study.

## I. INTRODUCTION

THE knowledge-based economy is based on the production, distribution and use of knowledge and information [1]. It is affected by the increasing use of information technologies to increase the competitive advantage in the economy. The main objective of the industry cluster development and supply chain management is to maintain competitiveness of the each industry in the market by using available information/knowledge. Though, industry cluster and supply chain are not the same aspect. Industry cluster is more in the macro-economic level which focuses on collaboration between partners in the same industry. But, Supply chain is more in the micro-economic level which focuses on the information sharing between companies who are in the same production chain. But, there are some focal points between two aspects which will be explained in the next section.

*A. Industry Cluster*

The concept of industry cluster was popularized by Prof. Michael E. Porter in his book named "Competitive Advantages of Nations" [2]. Then, industry cluster becomes current trend in economic development planning. Based-on Porter's definition about industry cluster [3], in this study the cluster is "geographically proximate group of companies and associated institutions (such as universities, government agencies, and related association) in a particular field, linked by commonalities and complementarities".

From the study of in 2005 [4] about the critical success factors in cluster development, first two critical success factors are collaboration in networking partnership and knowledge creation for innovative technology in the cluster which are about 78% and 74% of articles mentioned as success criteria. This knowledge is created through various forms of local inter-organizational collaborative interaction [5].

Another key success factor that supports the cluster development is Cluster Development Agent (CDA). CDA is a person who conceptualizes the overall developmental strategy for a cluster and initiates implementation. He is also the facilitator between the various cluster players and the target cluster. From statistic in cluster green book [6], 89% of successful clusters have full-time CDA.

*B. Knowledge Sharing and Collaboration in Cluster*

After the concept of industry cluster [2] was tangibly applied in many countries since 1998. Companies in the same industry trended to link to each other to maintain their competitiveness in the market and to gain benefits from being a member of the cluster. The major key success factors are knowledge sharing and collaboration within the cluster. This knowledge was collected in form of tacit and explicit knowledge in experts and organizations within the cluster.

Actually, knowledge is not a new idea [7] [8], as philosophers and scholars had been studying it for centuries. Due to the fact that the other three production factors (Land, Labor and Capital) were abundant, accessible and were considered the reason of economic advantage in the past, knowledge did not get much attention [9]. As tangible production factors are currently no longer enough to sustain a firm's competitive advantage, knowledge is being called on to play a key role [10].

*C. Knowledge Exchange in Supply Chain*

The collaboration in supply chain management (SCM) aims at increasing utilization and synchronization of the chain, resulting in tangible benefits for each participating company [11]. Within this context, the supply chain concept can be seen

P. Sureephong and N. Chakpitak are with College of Arts, Media and Technology, Chiang Mai University, Chiang Mai, Thailand. Email: dorn@camt.info, nopasit@camt.info.

L. Buzon and A. Bouras are with Laboratoire d'Informatique pour l'Entreprise et les Systèmes de Production (LIESP), University of Lyon 2, Lyon, France. Email: firstname.lastname@univ-lyon2.fr

as a collaborative network of organizations working together to maximize the value of a product to the final client. To be able to reduce conflict within this complex system, companies need to have common goals, clearly defined domains and especially a uniform understanding of situations.

The ability to make rapid decisions constitutes a competitive advantage [7] by decreasing the cycle time and by increasing the flexibility in order to respond to the change of customer's demand [12]. In this way, enterprises share knowledge to improve the global value carried out by the supply chain [13]. They share knowledge as product features, process information, best practices, allowed adjustments, dictionaries, etc., as illustrated in the figure 1, which describes example of knowledge sharing in the stool supply chain.

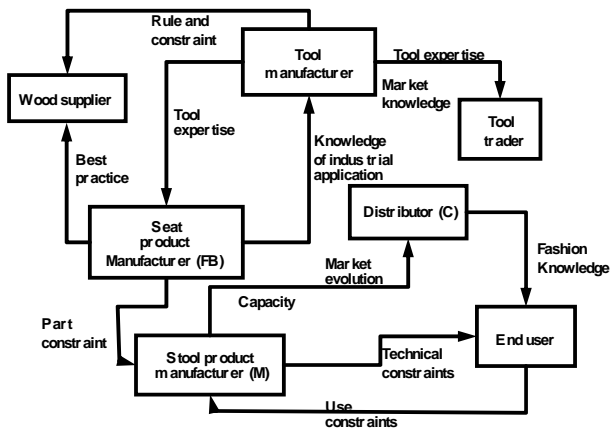

Fig. 1. Example of knowledge sharing in the stool supply chain.

We assume that the information, which is source of knowledge, circulates across the supply chain through each relationship between the partners. This means that each partner controls and interprets the flow of information according to their point of view and their understanding. Indeed, the information, received from one or several partners, are manipulated, analyzed and transmitted to the other partner in a synthetic form. Each firm, treats, retains information or favorites its exchange in keeping with the enterprise's strategy. So, we consider that the entire supply chain is composed by all the relationships between each partner of the supply chain [14]. The relationship is the fundamental of the knowledge sharing in the supply chain. All the relationships that participate to the production process can be aggregated to handle the diffusion and to measure the impact of knowledge sharing all along this process.

### D. Cluster and Supply Chain

The supply chain is integrated in the core business of the industry cluster. Both of them focus on improving the competitive advantage over their competitors. The cluster support the supply chain by integrating academic institutes, government agencies, association and supporting industry in order to create the innovation and enhance the knowledge in the supply chain. Figure 2 demonstrate the general view of the integration of industry cluster and supply chain.

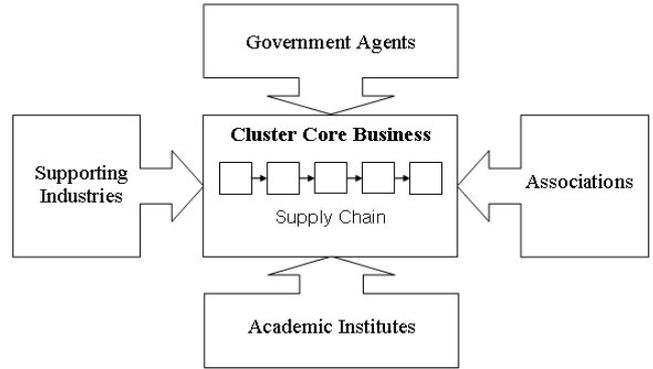

Fig. 2. Cluster map and supply chain.

As we described at the beginning that industry cluster and supply chain are in the different level of management but focus on the common objectives. However, there are many benefits of cluster to supply chain. In this paper, we focus on the benefit of the cluster in the supply chain core activities as follows,

--Clustering can improve the capability of companies to select their supply chain partner.

--Collaboration in the cluster will improve knowledge and information sharing between partners in the supply chain.

--CDA in the cluster can be Supply Chain Facilitator (SCF) which will assists the information sharing in the supply chain context.

## II. METHODOLOGY

This proposed methodology is based on supply chain management context by using the advantages of industry cluster to improve the collaboration and capability of exchanging knowledge. We proposed the methodology for supply chain to create relationship with new strategic partner. After that, the proposed framework of knowledge exchange will help supply chain to construct and classify their knowledge between actors.

### A. Knowledge Partner Management

The first step of collaboration in the supply chain management is searching and establishing relationship with the partner. The relationship is defined by two organizations which collaborate within the chain. Each partner controls and interprets the flow of information according his point of view. The information is manipulated, analyzed and transmitted to the other partner in a synthetic form. Thus, we consider that the entire supply chain is composed by all the relationship between each knowledge partner of the supply chain as shown in figure 3.

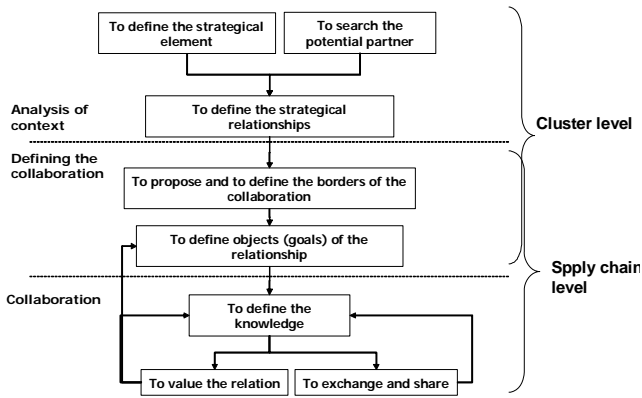

Fig. 3. The approach of the knowledge partner management.

The 'analysis of the context' step, firm has to select their strategic partner to collaborate with because it cannot afford to manage all relationship with supplier or customer because it is not worth enough. The collaboration in the cluster make possible for firms to find their potential partner which suite their strategic elements.

The 'defining the collaboration' step allows enclosing the relationship between borders and selecting the important issue between the partners. The relationship between firms in the cluster is transforming to a stronger relation in term of supply chain. Then, partners set their goals of the relationship together. In this step, cluster dimension aims to manage the relationship and supply chain dimension aim to handle the project notion and the goals of the relationship.

The 'collaboration' step focuses on knowledge defining and knowledge sharing. The knowledge which is shared in this step is mostly oriented to supply chain context. However, increasing in collaboration at the cluster level will help partners maintain knowledge sharing in supply chain level.

*B. CDA and SCF*

The CDA is one of key success factors of the cluster development. We can retain this concept to develop the notion of the Cluster Development Agent (CDA) as a Supply Chain Facilitator (SCF). Its role is to favor the collaboration in the supply chain by connecting the enterprise according their strategic goals and their capacities. In tactical perspective, the project in the relationship must be managed by independent person to preserve the unbiased negotiation. This means that the institute, such as universities or state organizations in the relationship is suitable to be SCF in the projects.

In knowledge exchange context which is the core principle of cluster and supply chain, SCF can help the actors to capture, construct or classify the knowledge by using knowledge card and the knowledge cube which are presented in the following section.

*C. Knowledge Elicitation*

The aim of the proposed model is to facilitate the communication and the exchange of knowledge between the actors of the supply chain. We use the notion of knowledge cards that is easy to handle in order to facilitate the user utilization (figure 4) [15].

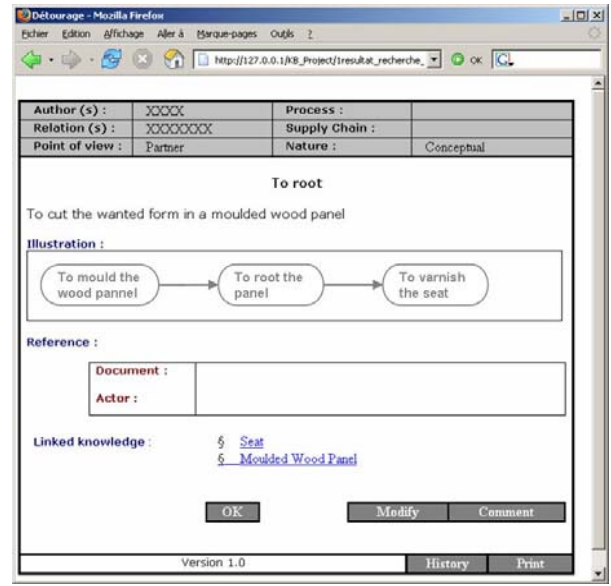

Fig. 4. Example of the translation of a knowledge card.

The header of the knowledge card represents the meta-data like the knowledge classification or the author. Then, the body of card is a textual description and illustration of the concept (the production processes). The references are elements that can help users to understand the knowledge card and also links to others knowledge cards that are complementary. The end of the knowledge card allows experts to modify or add comment to the knowledge system.

*D. Knowledge Classification*

The knowledge classification methodology is used to define and classify the exchanged and shared knowledge between partners. The proposed method to classify knowledge in supply chain is knowledge cube (figure 5). It helps user to search the knowledge elements when needed, in the organizational perspective and type in the project context. Knowledge cube can be divided in two sub-cubes, combining the explicit dimension that is linked to documents, repository, multimedia base, etc., and the tacit dimension that is linked to experts or group of experts as community of practices.

To create link between the knowledge, we use the object links (is-a, is a part of, is a kind of) and a regulation link that can be seen as an association link. This model is used to classify and structure the links between knowledge cards [15].

Fig. 5. The explicit knowledge classification.

*E. IT support Knowledge Exchange*

The objective of the information technology is to facilitate the communication and the exchange of knowledge between the actors of the supply chain within the cluster. It is very useful to help e-collaboration between the partner of the supply chain by reducing the cost and the delay of access to the information.

The knowledge cube represents the abstract level of the knowledge as an ontology domain, which provides a vocabulary of concepts. This ontology helps to define a generic conceptualization, shared by the supply chain actors. Ontology is used as unifying framework for people having different points of view. The approach is implemented via the Wiki concept. All the interactions and relationships appear in the knowledge card due to they can be modified or commented by the actors and all the document's history is saved as in the Wiki system.

### III. CASE STUDY

In our case study, we selected a supply chain of French stool which is in furniture cluster. This supply chain is composed of wood supplier, seat producer, stool producer, distributor and customer.

We constitute this example by assisting Chanel (the distributor: C), FB (the supplier: FB) and Mirima (the stool producer: M). In this case study, we will describe the Mirima's point of view in the relation of the supply chain.

*A. Analysis of the relation's context*

The clients of stool producer are luxury retailers who need very short delay time. In this context, M who is its supplier, searches knowledge elements that can help to judge, understand, control complex situations and resolve difficult problems with appropriate decision [17] in order to decrease the delivering time and propose innovation. M needs to know exactly when it has to order to receive the right product on the time.

In this step, the collaboration within the furniture cluster is very useful for M for searching a partner which suit defined knowledge elements. In this case, FB and M which are in the same cluster and used to collaborate with their partners in an informal manner, using phone and visits which concern suggestion about the form of the seat. There is a high level of confidence and trust between the two partners. They work together to install a varnishing facility in the FB production unit to reduce the delivery time.

*B. Knowledge Classification and Structuring*

From the knowledge cube, we found that the partners used to work together because there is less knowledge about concepts but more in mutual experience. Moreover, the reference of production process card shows the production activity of M. This reference is linked to the knowledge card that describes the M. Production process.

In the proposed methodology, knowledge card was used to represent the knowledge for knowledge users. We apply wiki concept with the knowledge card (figure 8), called Wiki Card, in order to structure knowledge in appropriate form. Each contribution can be read by all the members of a relationship and they can modify the knowledge if there is a mistake by editing the content.

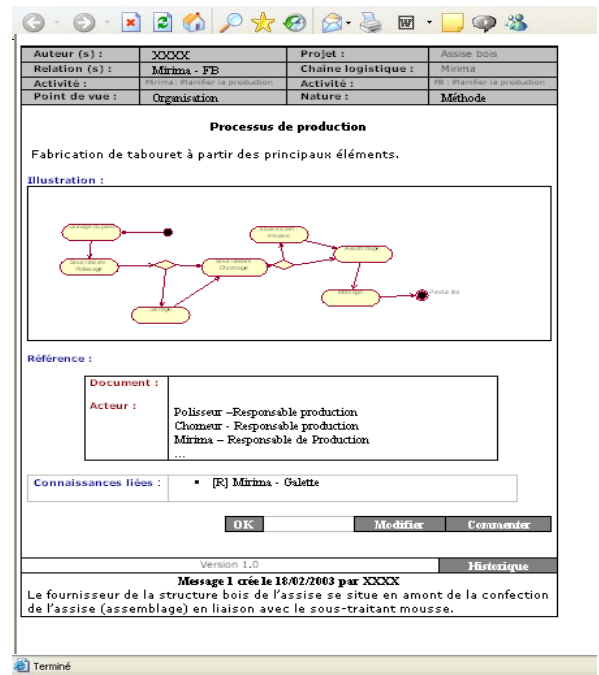

Fig. 6. An original knowledge card.

### IV. RESULT

In order to help users to reuse the knowledge, we associate the knowledge element with the activity of the process, in a workflow perspective (figure 7). Thus, this will help partners to improve their performance of their collaboration process. Furthermore, it helps to affine the contextualization of the knowledge element.

Like in the product flow or the information flow, the knowledge elements circulate in the supply chain through the relation between each actor. Members in the supply chain can control the knowledge elements that distribute to its partner.

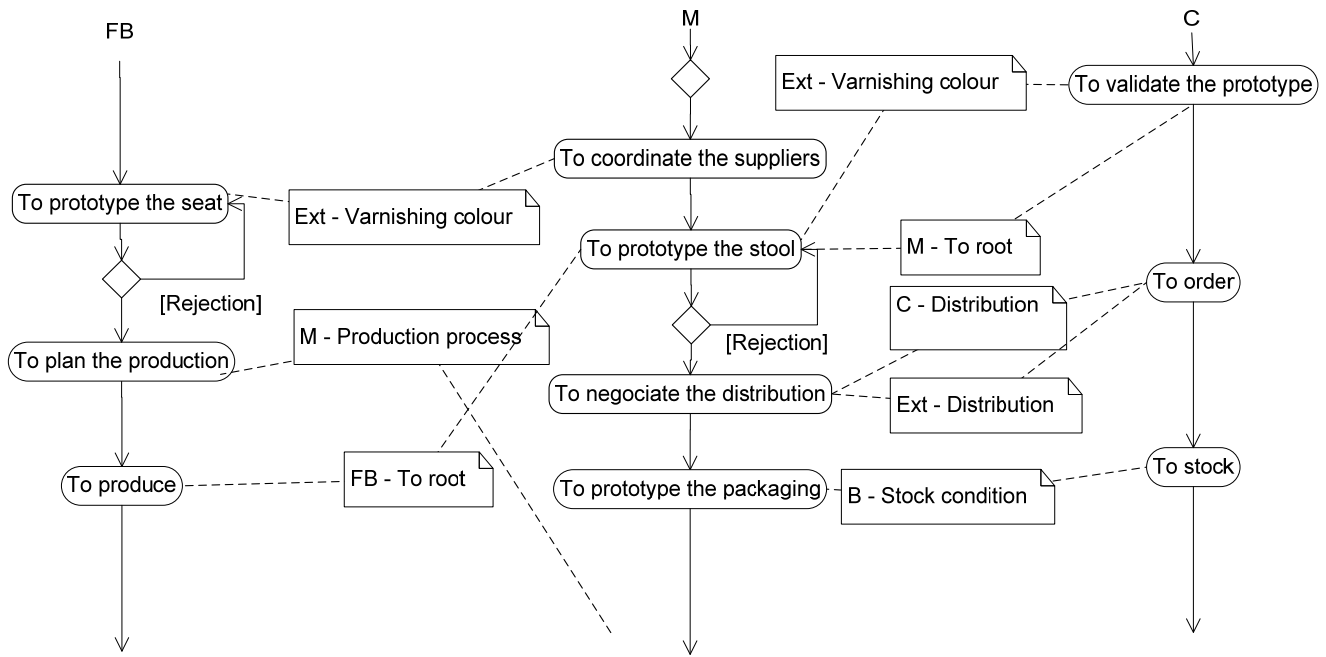

Fig. 7. Flow of knowledge exchange in supply chain.

FB and M exclaimed that this method reveals tacit and recurrent problems that need to be solved. It improves the confidence and reinforces the motivation to work together. Moreover, it reduced the risk to invest in new ideas and maximize opportunities in order to try new things. They explained that the good relationship depends on the relational capacity of their interlocutor.

This case of study shows that the informal interactions like the phone calls or visits are essential and the tool is only a means to have a better communication. First, they show difficulties to formalize knowledge and to choose which knowledge has to be formalized. Most of knowledge cards contained only text and reference to other documents because of the lack of time of the users. The users ask the possibility to import and export some knowledge cards to other projects with other partners that show their interest for this type of tools.

So far, the system was used by Mirima to communicate with its partners. The approach and methodology improve the effectiveness of Mirima in its capacity to meet Chanel's needs and develops the reactivity and the reliability of the FB and M relationship. Building of the knowledge cards was the pretext of discussion about important subject. The objective description of the relationship points out tacit problems that lead to negotiation and talk (creation of knowledge) and better understanding of each partner.

## V. CONCLUSION

This study focuses on the collaboration and knowledge exchange between partners in the supply chain context. We proposed the methodology for supply chain to create relationship with new strategic partner. We believe that industry cluster could improve supply chain performance. Conversely, improving knowledge exchanging in the supply chain would help cluster as well. The proposed framework of knowledge exchange between actors within a supply chain will help supply chain to construct and classify their knowledge. From the result of the study, these factors are important to supply chain and cluster for developing their collaboration.

-- *Trust and commitment*: The explicit implication in the project shows a will to become long term partners with more implication in the process of improving the collective benefits.

-- *Communication*: The explicit communication is improved but the informal (such as phone call) did not evolve significantly. The tool helps them to explicate their principle functional system that improves the whole system; they better understand the limits, constraints and priorities of each other.

-- *Adaptation*: Even if the price of the product has increased, they put the priority on the time of delivery. The quality of product sale increases and the answer to special demand is faster.

The future work will deal with the study of the knowledge echange in the from the design of product, manufacturing, …, to the waste valorisation of product in the cluster. This topic will focus on the sustenaible development of cluster to improve the global valor bring to the local population.

ACKNOWLEDGMENT

The authors would like to thank Euro-Asia Collaboration and NetWorking in Information Engineering System Technology (EAST-WEST) project which aim to increase the collaboration between European countries and Asian countries in term of economics, knowledge, innovation and human resource development.